# Synchronization of Two and Three Optoeletronic Oscillators Operating in Chaotic Regime: Numerical Simulations and Schemes for Secure Communication

[1,2] G. L. de Oliveira and [1,2] R. V. Ramos

glaucionor@gmail.com    rubens.viana@pq.cnpq.br

[1]*Federal Institute of Education, Science and Technology of Ceara, Fortaleza-Ce, Brazil.*
[2]*Laboratory of Quantum Information Technology, Department of Teleinformatic Engineering – Federal University of Ceara - DETI/UFC, C.P. 6007 – Campus do Pici - 60455-970 Fortaleza-Ce, Brazil.*

In this work we show a strategy for synchronization of three optoelectronic oscillators (OEO) operating in chaotic regime. Two applications of synchronized OEOs in secure communications are considered. In the first one the OEO is used to produce a pseudo-random bit sequence. The second application is an optical setup for secure transmission of sampled analog signals. Using numerical simulations, we calculated the bit error rate taking into account parameter mismatch noise and Gaussian noise in the input optical power. The conditions for error rate of up to 15% during key generation are shown.

## 1. Introduction

The rich dynamic shown by non-linear systems operating in chaotic regime has found applications in several areas. An important example is the encryption/decryption of messages for secure communication between distant authorized parts. In optical systems the chaotic behavior has been studied, for example, in optically injected lasers [4,5], nonlinear optical resonator, where the Kerr effect is used [6-8], and optoelectronic oscillators (OEO) [9]. For the last, the nonlinearity in the feedback line is obtained by the detection of the light: the photocurrent is proportional to the impinging optical power and this behavior does not depend on the value of the incident optical power. In this work we consider exclusively the chaotic behavior in the OEO described in [10]. This OEO produces chaotic light polarization states.

In order to implement a secure communication system using the properties of chaotic systems (pseudo-randomness and high dependence on the parameters values), two or more chaotic systems must be synchronized. The synchronization of chaotic systems has been firstly addressed in [8,9] and the realization of secure optical system employing synchronized chaotic systems have been reported in the literature [13-21].

In this work, we show numerically the synchronization of two and three OEOs operating in chaotic regime, as well its application in digital and analog secure communication systems. In the first case, each synchronized chaotic system (one in Alice, the transmitter, and the other in Bob, the receiver) generates a binary key from the quantization of the Stokes' parameter $S_1$ of the chaotic output light polarization state. In a second step, that key is used for secure communication through one-time pad protocol, for example. Furthermore, the error rate in the binary key is calculated numerically with and without noise in the parameters values. In what concerns the analog communication system, we present a setup for secure communication of sampled analog signals.

This work is outlined as follows: In Section II we describe the OEO. In Section III it is shown the synchronization of two and three OEOs operating in chaotic regime. The digital and analog secure communication systems are described in Section IV. At last, the conclusions are drawn in Section V.

## 2. Generation of Chaotic Polarization States

Basically the optoelectronic oscillator is a system where the light emitted by the laser is modulated (by an electrooptical modulator) and detected. The photocurrent produced is amplified and used to feed the laser itself or the electrooptical modulator. The optoelectronic oscillator considered in this work is the one that produces chaotic polarization states described in [10]. It is shown in Fig. 1.

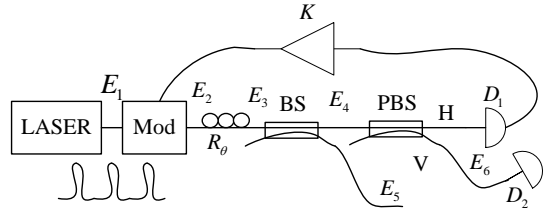

Fig. 1 – Optoelectronic oscillator for generation of chaotic polarization states: BS – beam splitter, PBS – Polarizing beam splitter, $K$ – Electrical amplifier, $D_{1,2}$ – Photodetectors, $R_\theta$ - Polarization rotator, MOD – electrooptical polarization modulator and $E_1$-$E_6$ are the electrical fields at the marked positions.

In Fig. 1, the time required to the light be detected in $D_1$ and the electrical signal produced to feed the optical modulator (Mod) is equal to the time interval between two consecutive pulses generated by the laser. BS is a balanced optical coupler, PBS is a polarizing beam splitter, $D_1$ and $D_2$ are optical detectors and $K$ is an electrical amplifier.

The equations that describe the polarization states in Fig. 2 are [22]

$$E_1 = |\alpha, \alpha\rangle_{HV} \tag{1}$$

$$E_2 = \left|\alpha \exp\left(j\left(\frac{\pi}{2}\frac{V_{in}(t)}{V_\pi}+\varphi\right)\right), \alpha \exp\left(-j\left(\frac{\pi}{2}\frac{V_{in}(t)}{V_\pi}+\varphi\right)\right)\right\rangle_{HV} \tag{2}$$

$$E_3 = \left|i\sqrt{2}\alpha \sin\left(\frac{\pi}{2}\frac{V_{in}(t)}{V_\pi}+\varphi\right), \sqrt{2}\alpha \cos\left(\frac{\pi}{2}\frac{V_{in}(t)}{V_\pi}+\varphi\right)\right\rangle_{HV} \tag{3}$$

$$E_4 = \left|i\alpha \sin\left(\frac{\pi}{2}\frac{V_{in}(t)}{V_\pi}+\varphi\right), \alpha \cos\left(\frac{\pi}{2}\frac{V_{in}(t)}{V_\pi}+\varphi\right)\right\rangle_{HV} \tag{4}$$

$$E_5 = \left| -\alpha \sin\left(\frac{\pi}{2}\frac{V_{in}(t)}{V_\pi} + \varphi\right), i\alpha \cos\left(\frac{\pi}{2}\frac{V_{in}(t)}{V_\pi} + \varphi\right) \right\rangle_{HV} \qquad (5)$$

$$E_6 = \left| 0, i\alpha \cos\left(\frac{\pi}{2}\frac{V_{in}(t)}{V_\pi} + \varphi\right) \right\rangle_{HV} \qquad (6)$$

$$V_{in}(t+\tau) = K|\alpha|^2 \sin^2\left(\frac{\pi}{2}\frac{V_{in}(t)}{V_\pi} + \varphi\right). \qquad (7)$$

In (1)-(6) the Dirac notation $E = |E_x, E_y\rangle_{HV}$ corresponds to the electrical field column vector $[E_x\ E_y]^T$ and the subscripts $H$ and $V$ mean, respectively, horizontal ($E_x$) and vertical ($E_y$) modes. In (7) $\tau$ is the time interval between two consecutives light pulses. The light produce by the laser is linearly polarized in $\pi/4$. The optical modulator adds a phase of $\pi V_{in}/(2V_\pi)+\varphi$ in the horizontal component and $-\pi V_{in}/(2V_\pi)-\varphi$ in the vertical component. Here, $V_{in}$ is the modulating signal, $V_\pi$ is the voltage required to add a $\pi/2$ phase and $\varphi$ is the offset value. The polarization rotator applies a $\pi/4$ rotation in the input state. After the polarization rotator, a beam splitter is used to divide the optical signal. One half is the output state, $E_5$, and the other half has its horizontal and vertical components separated by a polarizing beam splitter. The horizontal part is detected in $D_1$ and the resulting photocurrent is amplified and used as modulating signal. The vertical component is detected in $D_2$ and its value is used for synchronization purposes as it will be shown latter.

The chaotic behavior appears for appropriate values of $K$, that models the electrical amplifier gain, optical losses and the detector's ($D_1$) efficiency, the optical input power ($|\alpha|^2$), $\varphi$ and $V_\pi$. The initial value of $V_{in}$ depends on the internal noise of the electronic devices. At last, the recurrence equation that shows the feedback is (7). For suitable parameters values, the variable $V_{in}$ shows a chaotic behavior, as shown in Fig. 2: The upper part is the bifurcation diagram versus $|\alpha|^2$ while the lower part shows the Lyapunov number $\lambda$ versus $|\alpha|^2$. The chaotic regime is obtained for values $|\alpha|^2$ for which $\lambda > 0$ ($\varphi = \pi/4$, $V_\pi = 1$, $K = 0.0133$ and $V_{in}(0) = 0.1$).

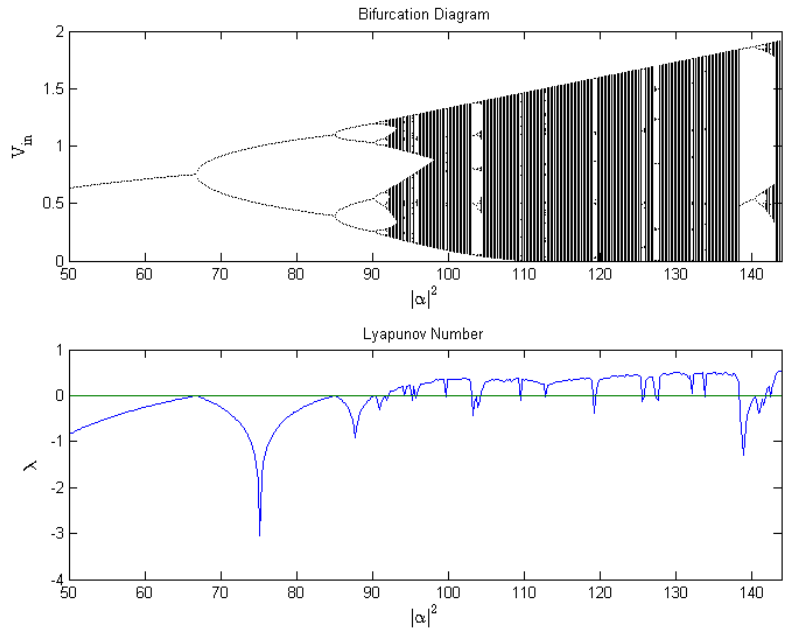

Fig. 2 – Bifurcation diagram and Lyapunov number ($\lambda$) versus $|\alpha|^2$ for equation (7).

Since $V_{in}$ shows a chaotic behavior the polarization state of $E_5$ is chaotic too.

## 3. Synchronization of Two and Three OEOs Operating in Chaotic Regime

The usage of chaotic systems in secure communication tasks requires the synchronization between two or more chaotic systems. To maintain that synchronism is not a trivial task, mainly in the presence of noises in the initial conditions and parameters values. In order to keep the synchronism, the chaotic systems have to exchange information. The synchronization of two chaotic systems of the type shown in Fig. 1 is as shown in Fig. 3 [22].

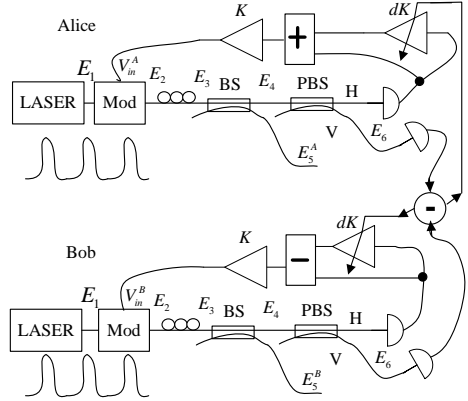

Fig. 3 – Two synchronized OEOs producing chaotic light polarization states.

As it can be seen in Fig. 3, in order to get the synchronization, information about the optical power of the vertical output $E_6$ is sent from one system to the other. The difference between them is the control variable used to keep the two chaotic systems synchronized and the controlled variable is the amplifier gain $dK$. The correction for both systems is

$$dK = \delta |\alpha|^2 \left[ \cos^2\left( \frac{\pi}{2} \frac{V_{in}^B(t)}{V_\pi} + \varphi \right) - \cos^2\left( \frac{\pi}{2} \frac{V_{in}^A(t)}{V_\pi} + \varphi \right) \right]. \tag{8}$$

In (8), $\delta$ is a constant related to the optical detector. The controlled variables are $V_{in}^A$ (+) and $V_{in}^B$ (-):

$$V_{in}^{A,B}(t+\tau) = K|\alpha|^2 \sin^2\left( \frac{\pi}{2} \frac{V_{in}^{A,B}(t)(1 \pm dK)}{V_\pi} + \varphi \right). \tag{9}$$

Figure 4 shows the bifurcation diagram of the Stokes' parameter $S_1$ of $E_5$ (the difference between the optical power in the horizontal and vertical modes).

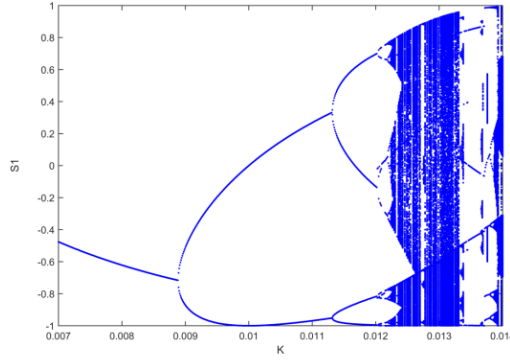

Fig. 4 - Bifurcation diagram for the synchronized systems shown in Fig. 3.

The value of $S_1$ is given by

$$S_1 = \varepsilon\left[\left|-\alpha\sin\left(\frac{\pi}{2}\frac{V_{in}}{V_\pi}+\varphi\right)\right|^2 - \left|i\alpha\cos\left(\frac{\pi}{2}\frac{V_{in}}{V_\pi}+\varphi\right)\right|^2\right] = -\varepsilon|\alpha|^2\cos(\pi V_{in}/V_\pi + 2\varphi), \tag{10}$$

where $\varepsilon$ is a constant that takes into account details of the polarimeter used to measure $S_1$. In this simulation the following parameters values were used: $\varepsilon = 0.01$, $|\alpha|^2 = 100$, $\varphi = \pi/4$, $V_\pi = 1$V, $\delta = 0.015$, $K \in [0.007, 0.014]$, $V_{in}^A(t=0) = 0.1$ and $V_{in}^B(t=0) = 0.2$ (the two chaotic systems are equal but they start with different initial conditions).

Using the same parameters values described before and $K = 0.0132$, the result of the synchronism can be seen in Figs. 5 and 6. Figure 5 shows $\Delta = V_{in}^A(t) - V_{in}^B(t)$ during the first 4000 pulses. The synchronization strategy is turned on only after the 100$^{th}$ optical pulse. As it can be seen in Fig. 5, the dynamics changes from almost perfect to non-perfect (but still good) synchronism several times. An example of this transition is shown in Fig. 6.

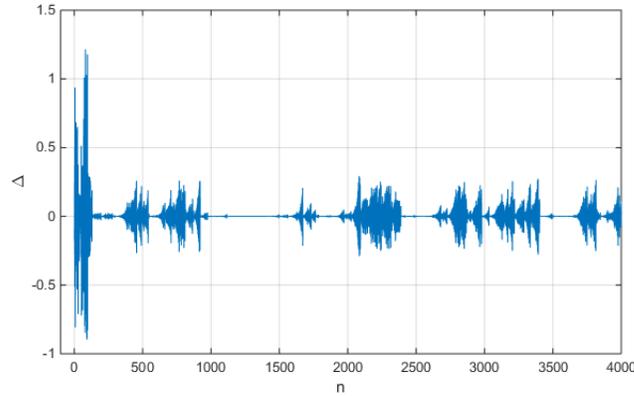

Fig. 5 - Result of the synchronization of two OEOs in chaotic regime during the first 4000 pulses: $\Delta = V_{in}^A - V_{in}^B$.

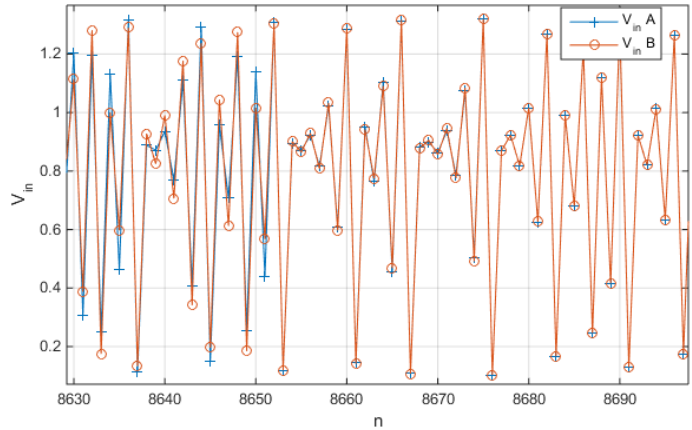

Fig. 6 - $V_{in}^A$ and $V_{in}^B$ versus round trip number: Transition from non-perfect synchronism to almost perfect synchronism.

The desynchronization shown in Figs. 5 and 6 is due to the parameter mismatch noise [23]. It can happen in any kind of synchronized chaotic systems. In our case, the values of $V_{in}^A(t = 0)$ and $V_{in}^B(t = 0)$ are, intentionally, made different. For the result shown in Fig. 4 we used $V_{in}^A(t = 0) = 0.1$ and $V_{in}^B(t = 0) = 0.2$. If we had done $V_{in}^B(t = 0) = V_{in}^A(t = 0)$ the mismatch noise would not exist and the desynchronization would disappear. On the other hand, if one increases the mismatch ($V_{in}^B(t = 0) \gg V_{in}^A(t = 0)$ for example), the desynchronization increases in two ways: the amplitude of $\Delta$ increases and the duration of synchronization intervals decrease.

The natural next step is the synchronization of three OEOs working in the chaotic regime. Figure 7 shows the scheme to synchronize three OEOs.

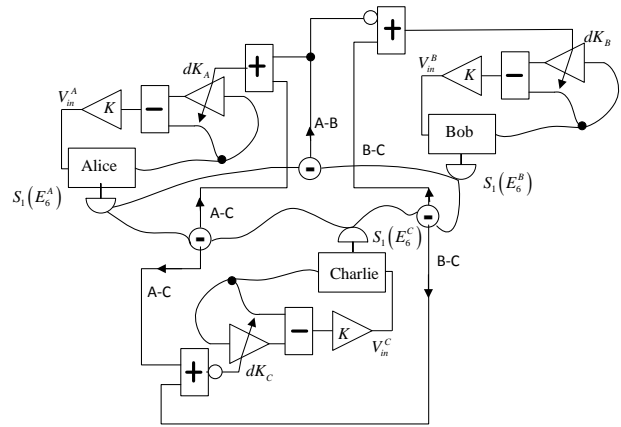

Fig. 7 – Scheme for synchronization of three OEOs producing chaotic light polarization states. The white balls are inverters.

The scheme shown in Fig. 7 is a natural extension of the scheme in Fig. 3. The modulating signals are

$$V_{in}^{A}(t+\tau) = K|\alpha|^{2} \sin^{2}\left(\frac{\pi}{2}\frac{V_{in}^{A}(t)(1-dK_{A})}{V_{\pi}}+\varphi\right) \tag{11}$$

$$V_{in}^{B}(t+\tau) = K|\alpha|^{2} \sin^{2}\left(\frac{\pi}{2}\frac{V_{in}^{B}(t)(1-dK_{B})}{V_{\pi}}+\varphi\right) \tag{12}$$

$$V_{in}^{C}(t+\tau) = K|\alpha|^{2} \sin^{2}\left(\frac{\pi}{2}\frac{V_{in}^{C}(t)(1-dK_{C})}{V_{\pi}}+\varphi\right), \tag{13}$$

where the corrections are given by

$$dK_{A} = \delta_{A}|\alpha|^{2}\left\{2\cos^{2}\left(\frac{\pi}{2}\frac{V_{in}^{A}}{V_{\pi}}+\varphi\right)-\cos^{2}\left(\frac{\pi}{2}\frac{V_{in}^{B}}{V_{\pi}}+\varphi\right)-\cos^{2}\left(\frac{\pi}{2}\frac{V_{in}^{C}}{V_{\pi}}+\varphi\right)\right\} \tag{14}$$

$$dK_{B} = \delta_{B}|\alpha|^{2}\left\{2\cos^{2}\left(\frac{\pi}{2}\frac{V_{in}^{B}}{V_{\pi}}+\varphi\right)-\cos^{2}\left(\frac{\pi}{2}\frac{V_{in}^{A}}{V_{\pi}}+\varphi\right)-\cos^{2}\left(\frac{\pi}{2}\frac{V_{in}^{C}}{V_{\pi}}+\varphi\right)\right\} \tag{15}$$

$$dK_{C} = \delta_{C}|\alpha|^{2}\left\{2\cos^{2}\left(\frac{\pi}{2}\frac{V_{in}^{C}}{V_{\pi}}+\varphi\right)-\cos^{2}\left(\frac{\pi}{2}\frac{V_{in}^{A}}{V_{\pi}}+\varphi\right)-\cos^{2}\left(\frac{\pi}{2}\frac{V_{in}^{B}}{V_{\pi}}+\varphi\right)\right\}. \tag{16}$$

The results of a simulation with 10,000 pulses of the synchronization of three OEOs operating in chaotic regime can be seen in Figs. 8 and 9. The parameters values are $|\alpha|^2 = 100$, $\varphi = \pi/4$, $V_\pi = 1\text{V}$, $\delta_A = 0.0010$, $\delta_B = 0.006$, $\delta_C = 0.0013$, $K = 0.0132$, $V_{in}^A(t = 0) = 0.1$, $V_{in}^B(t = 0) = 0.2$ and $V_{in}^C(t = 0) = 0.15$ (the three OEOs are equal but they start with a different initial condition). Figure 8 shows $V_{in}^A - V_{in}^B$, $V_{in}^A - V_{in}^C$ and $V_{in}^B - V_{in}^C$ during the first 2000 pulses. The synchronization strategy is turned on only after the 100$^{\text{th}}$ optical pulse. From Fig. 8 one can note that OEO$_A$ and OEO$_C$ keep well synchronized. Figure 9, by its turn, shows a transition from non-perfect to quasi-perfect synchronization.

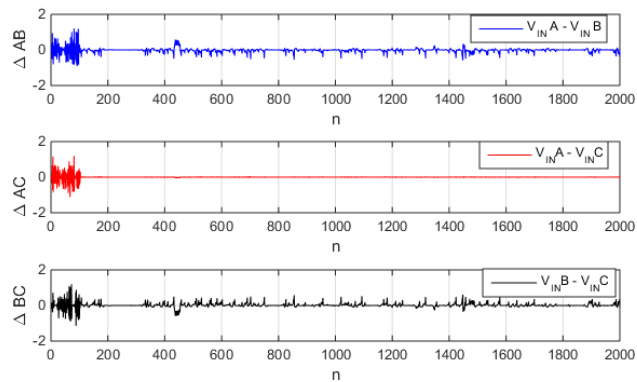

Fig. 8 - Result of the synchronization of three OEOs operating in chaotic regime: $\Delta AB = V_{in}^A - V_{in}^B$, $\Delta AC = V_{in}^A - V_{in}^C$ and $\Delta BC = V_{in}^B - V_{in}^C$.

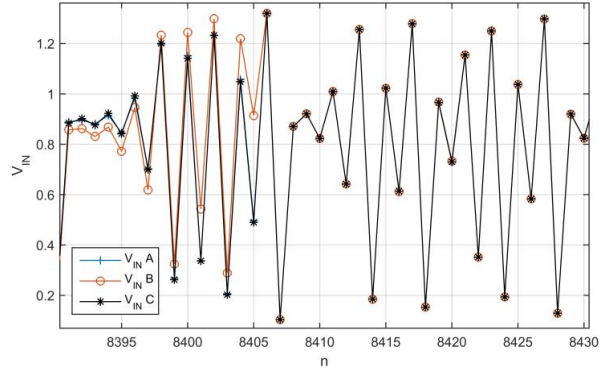

Fig. 9 - $V_{in}^A$, $V_{in}^B$ and $V_{in}^C$ versus round trip number: Transition from non-perfect synchronism to almost perfect synchronism.

The very good synchronization between OEO$_A$ and OEO$_C$, when compared to OEO$_B$, happens because, for the set of parameters values used, the mismatch between $\delta_A$ and $\delta_C$ is much smaller than the mismatch between any one of them and $\delta_B$: $|\delta_A - \delta_C| \ll \min(|\delta_A - \delta_B|, |\delta_B - \delta_C|)$. Exchanging the values of $\delta_B$ and $\delta_C$ and keeping all the other parameters with the same values, would yield a very good synchronization between OEO$_A$ and OEO$_B$. At last, if we had used, for example, $\delta_A = \delta_B = \delta_C = 0.0013$, the three OEOs would be very well synchronized however in a periodic regime.

## 4. Aplication of Synchronized Chaotic OEOs in Secure Communication Systems

In the simplest secure communication system, two authorized users, Alice and Bob, share a common binary key, $X^k$, usually generated by a pseudorandom number generator (PRNG). A secure communication is achieved if Alice sends to Bob the encoded message $X^k \oplus M$ where $M$ is the secret message having the same number of bits of the key. Furthermore, the key must be used only once. Since Bob has the same key, he can recover the message by doing $X^k \oplus (X^k \oplus M) = M$. In this direction, the most obvious usage of chaotic systems in secure communication is to use them as PRNGs. That is, each chaotic system (one in Alice and the other in Bob) plays the role of the PRNG. Since they are synchronized, one expects them to generate the same key in Alice and Bob. When Alice and Bob use computer-based PRNG, they must use the same seed (another binary sequence shared by Alice and Bob in advance) in order to produce the same binary strings that will be used as key. When chaotic systems are used as PRNG, the initial secret corresponds to the parameters of the non-linear system. Similar non-linear systems with different parameters values will show different dynamics. The output $X^k$ is the bit sequence formed by the discretization of the Stokes' parameter $S_1$ of the output optical field $E_5$ given in (10). In order to obtain a binary sequence from the continuous values of $S_1$, a threshold value $S_{th}$ is defined and, when $S_1 < S_{th}$ the bit '0' is obtained otherwise the bit '1' is obtained. For the example given in Figs. 5 and 6, using $\varepsilon = 0.01$ in (10) and $S_{th} = 0.7$, one gets a bit error rate (BER) between the bit sequences obtained by Alice and Bob of 4.78% in 10,000 bits. This error is due to the non-perfect synchronism regions as those shown in Fig. 5. In practice, an error correction protocol can be used and perfect correlation between the bit sequences, $X^k(V_{in}^A)$ obtained by Alice and $X^k(V_{in}^B)$ obtained by Bob can be achieved.

In order to analyze the influence of noise in the error rate between Alice and Bob raw keys (before error correction) we calculated numerically the BER when a Gaussian noise is added in Bob's amplifier's gain. In this case, Bob's modulating signal is

$$V_{in}^B(t+\tau) = K\left(1+\frac{x(t)}{N}\right)|\alpha|^2 \sin^2\left(\frac{\pi}{2}\frac{V_{in}^B(t)(1-dK)}{V_\pi}+\varphi\right). \tag{17}$$

In (17) the variable $x$ is a normally distributed random variable with mean zero and standard deviation equal to 1. The variable $N$ controls the strength of the noise. Varying $N$ in the range 10-10000, one can see in Fig. 10 the (average) bit error rate. The maximal bit error rate is 14.46% ($N = 10$) that is a value that still permits a reconciliation between Alice and Bob keys through an error correction protocol.

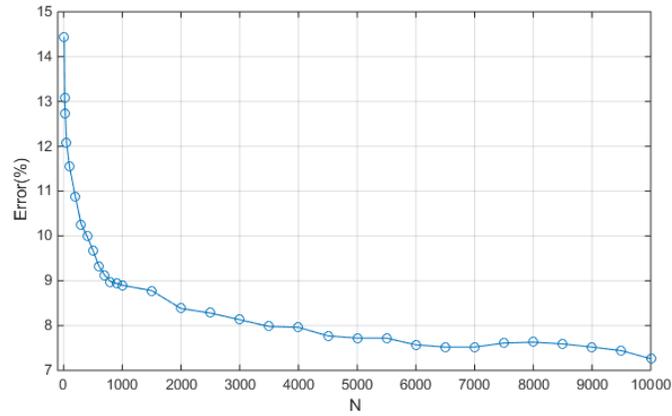

Fig. 10 - Average bit error rate versus noise strength $N$.

The usage of three OEOs in secure communications is a straightforward extension of what was just discussed. However, in this case the final key is shared between three authorized users: Alice, Bob and Charlie. Simulating a noiseless system (the BER is only due to the imperfect synchronism) with $\varepsilon = 0.01$ and $S_{th} = 0.72$, we got a BER of 10.73% between Alice and Bob, 10.63% between Alice and Charlie and 1.08% between Bob and Charlie (10,000 bits).

In order to analyze the error rate in the presence of noise, we plotted the average BER versus the strength of the noise in three situations: I) Noise only in Bob's setup. II) Noise only in Charlie setup. III) Noise in both Bob's and Charlie's setups. The equations for Bob and Charlie in the presence of noise are

$$V_{in}^B = K\left(1+\frac{x_B}{N_B}\right)|\alpha|^2 \sin^2\left(\frac{\pi}{2}\frac{V_{in}^B(1-dK_B)}{V_\pi}+\varphi\right) \tag{18}$$

$$V_{in}^C = K\left(1+\frac{x_C}{N_C}\right)|\alpha|^2 \sin^2\left(\frac{\pi}{2}\frac{V_{in}^C(1-dK_C)}{V_\pi}+\varphi\right). \tag{19}$$

As before, in (18)-(19) $x_B$ and $x_C$ are normally distributed random variables with mean zero and standard deviation equal to 1. The variables $N_B$ and $N_C$ control the strength of the noise in Bob and Charlie, respectively. The curves are shown in Fig. 11.

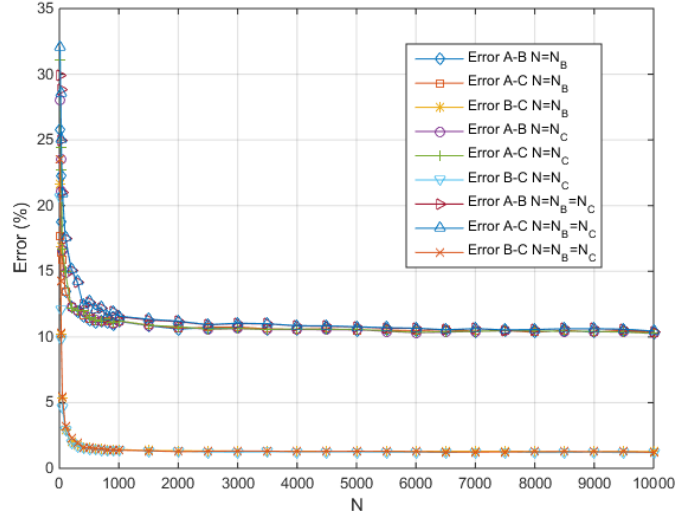

Fig. 11 – Bit error rate between the bit sequences obtained by synchronization of three OEOs operating in the chaotic regime versus the parameter $N$ in (18)-(19) ($N = N_B$ when there is noise in Bob and $N = N_C$ when there is noise in Charlie).

In the simplest error correction procedure, initially Alice and Bob divide their binary sequences in two parts, $K_A = [K_{A1}\ K_{A2}]$ and $K_B = [K_{B1}\ K_{B2}]$. They compare the parity of each part. If, for example, PAR($K_{A1}$) ≠ PAR($K_{B1}$), where PAR($X$) gives the parity of the number of bits '1' in $X$, then one knows an error occurred. In this case, $K_{A1}$ and $K_{B1}$ are divided in two subsets and a new parity comparison is applied in both subsets. The process is repeated until the final subsets with different parities have only two bits. Since one of them is wrong, both are discarded and none information is leaked. The new keys are scrambled and the whole process is repeated a number of times. The action of discarding a pair of bits implies in a decrease of the key's length. In average, the fraction of bits lost during error correction, $R_{ec}$, is given by $R_{ec}$ = (7/2)BER-BERlog$_2$(BER) [24]. Hence, in practice, the maximal error rate acceptable is around 15% ($R_{ec}$~0.935). Thus, according to the numerical results shown in Figs. 10 and 11, even in the presence of some external noise, a useful key can be obtained. Howerver, the system with three synchronized OEOs is more susceptible to noise.

Another possible application of two synchronized OEOs is shown in Fig. 12. It is a setup for secure transmission of sampled analog signals. Basically, the chaotic polarization state produced by Alice's OEO, $E_5^A$, has its polarization components separated by a polarizing beam splitter. The horizontal (vertical) component is phase-modulated by $\phi_H$ ($\phi_V$). Additionally, the horizontal component is time delayed of $\tau$. Thus, the vertical and horizontal components are launched in the optical channel in different times. When they arrive at Bob, they will suffer interference in the beam splitters with the horizontal and vertical components of $E_5^B$. If the OEOs are well synchronized one has $E_5^A \approx E_5^B$ and the results of the interferences will depend only on $\phi_H$ and $\phi_V$.

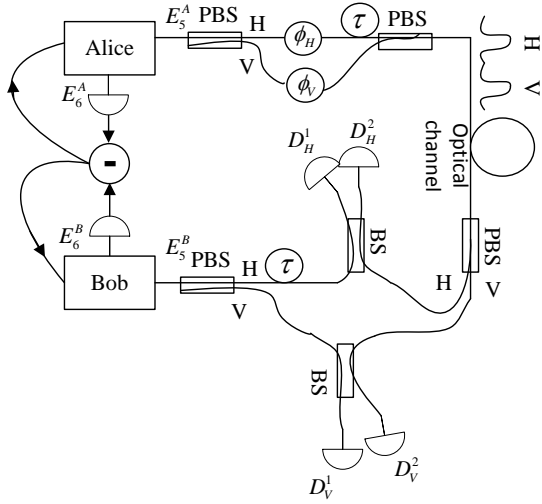

Fig. 12 – Optical setup for secure transmission of sampled analog signals employing chaotic polarization states.

Considering the balanced beam splitters action as

$$U_{BS}|\xi,\lambda\rangle = \left|\frac{\xi+\lambda}{\sqrt{2}}, \frac{-\xi+\lambda}{\sqrt{2}}\right\rangle, \quad (20)$$

the equations that explain the functioning of the setup in Fig. 12 are the following:

At the channel's input:

$$\left|-\alpha\sin\left(\frac{\pi}{2}\frac{V_{in}}{V_\pi}+\varphi\right)e^{i\phi_H}, i\alpha\cos\left(\frac{\pi}{2}\frac{V_{in}}{V_\pi}+\varphi\right)e^{i\phi_V}\right\rangle_{HV}. \quad (21)$$

At Bob's beam splitters' inputs:

$$\left|-\alpha\sin\left(\frac{\pi}{2}\frac{V_{in}^A}{V_\pi}+\varphi\right)e^{i\phi_H}, -\alpha\sin\left(\frac{\pi}{2}\frac{V_{in}^B}{V_\pi}+\varphi\right)\right\rangle \quad (22)$$

$$\left|i\alpha\cos\left(\frac{\pi}{2}\frac{V_{in}^A}{V_\pi}+\varphi\right)e^{i\phi_V}, i\alpha\cos\left(\frac{\pi}{2}\frac{V_{in}^B}{V_\pi}+\varphi\right)\right\rangle. \quad (23)$$

At Bob's beam splitters outputs

$$\left|\frac{-\alpha}{\sqrt{2}}\left[\sin\left(\frac{\pi}{2}\frac{V_{in}^{A}}{V_{\pi}}+\varphi\right)e^{i\phi_{H}}+\sin\left(\frac{\pi}{2}\frac{V_{in}^{B}}{V_{\pi}}+\varphi\right)\right],\right\rangle \tag{24}$$

$$\left|\frac{-\alpha}{\sqrt{2}}\left[-\sin\left(\frac{\pi}{2}\frac{V_{in}^{A}}{V_{\pi}}+\varphi\right)e^{i\phi_{H}}+\sin\left(\frac{\pi}{2}\frac{V_{in}^{B}}{V_{\pi}}+\varphi\right)\right]\right\rangle$$

$$\left|i\frac{\alpha}{\sqrt{2}}\left[\cos\left(\frac{\pi}{2}\frac{V_{in}^{A}}{V_{\pi}}+\varphi\right)e^{i\phi_{V}}+\cos\left(\frac{\pi}{2}\frac{V_{in}^{B}}{V_{\pi}}+\varphi\right)\right],\right\rangle \tag{25}$$

$$\left|i\frac{\alpha}{\sqrt{2}}\left[-\cos\left(\frac{\pi}{2}\frac{V_{in}^{A}}{V_{\pi}}+\varphi\right)e^{i\phi_{V}}+\cos\left(\frac{\pi}{2}\frac{V_{in}^{B}}{V_{\pi}}+\varphi\right)\right]\right\rangle.$$

The photocurrent produce in the detectors by the four fields in (24)-(25) are

$$I_{1}=R\frac{|\alpha|^{2}}{2}\left[\begin{array}{c}\sin^{2}\left(\frac{\pi}{2}\frac{V_{in}^{A}}{V_{\pi}}+\varphi\right)+\sin^{2}\left(\frac{\pi}{2}\frac{V_{in}^{B}}{V_{\pi}}+\varphi\right)+\\ 2\sin\left(\frac{\pi}{2}\frac{V_{in}^{A}}{V_{\pi}}+\varphi\right)\sin\left(\frac{\pi}{2}\frac{V_{in}^{B}}{V_{\pi}}+\varphi\right)\cos\left(\phi_{H}\right)\end{array}\right] \tag{26}$$

$$I_{2}=R\frac{|\alpha|^{2}}{2}\left[\begin{array}{c}\sin^{2}\left(\frac{\pi}{2}\frac{V_{in}^{A}}{V_{\pi}}+\varphi\right)+\sin^{2}\left(\frac{\pi}{2}\frac{V_{in}^{B}}{V_{\pi}}+\varphi\right)-\\ 2\sin\left(\frac{\pi}{2}\frac{V_{in}^{A}}{V_{\pi}}+\varphi\right)\sin\left(\frac{\pi}{2}\frac{V_{in}^{B}}{V_{\pi}}+\varphi\right)\cos\left(\phi_{H}\right)\end{array}\right] \tag{27}$$

$$I_{3}=R\frac{|\alpha|^{2}}{2}\left[\begin{array}{c}\cos^{2}\left(\frac{\pi}{2}\frac{V_{in}^{A}}{V_{\pi}}+\varphi\right)+\cos^{2}\left(\frac{\pi}{2}\frac{V_{in}^{B}}{V_{\pi}}+\varphi\right)+\\ 2\cos\left(\frac{\pi}{2}\frac{V_{in}^{A}}{V_{\pi}}+\varphi\right)\cos\left(\frac{\pi}{2}\frac{V_{in}^{B}}{V_{\pi}}+\varphi\right)\cos\left(\phi_{V}\right)\end{array}\right] \tag{28}$$

$$I_{4}=R\frac{|\alpha|^{2}}{2}\left[\begin{array}{c}\cos^{2}\left(\frac{\pi}{2}\frac{V_{in}^{A}}{V_{\pi}}+\varphi\right)+\cos^{2}\left(\frac{\pi}{2}\frac{V_{in}^{B}}{V_{\pi}}+\varphi\right)-\\ 2\cos\left(\frac{\pi}{2}\frac{V_{in}^{A}}{V_{\pi}}+\varphi\right)\cos\left(\frac{\pi}{2}\frac{V_{in}^{B}}{V_{\pi}}+\varphi\right)\cos\left(\phi_{V}\right)\end{array}\right]. \tag{29}$$

In (26)-(29) $R$ is the responsivity of the detectors. Now, for $\phi_H = \phi_V = \phi$, one has

$$I=(I_{1}-I_{2})+(I_{3}-I_{4})=R\frac{|\alpha|^{2}}{2}\cos(\phi)\cos\left(\frac{\pi}{2}\frac{V_{in}^{A}-V_{in}^{B}}{V_{\pi}}\right). \tag{30}$$

Here we will assume a sampled sinusoidal modulating signal $\phi = mA\sin(\omega kT)$, where $m$ is the modulation index, $k$ is an integer number and $T$ is the time step, equal to the time separation between consecutive optical pulses generated by the laser. For a small value of $m$ one has $\cos(\phi) \sim 1+\phi = 1+mA\sin(\omega kT)$. Substituting in (30) one finally gets

$$I=R\frac{|\alpha|^{2}}{2}\left[1+mA\sin(\omega kT)\right]\cos\left(\frac{\pi}{2}\frac{V_{in}^{A}-V_{in}^{B}}{V_{\pi}}\right). \tag{31}$$

If the synchronism between the OEOs is perfect, then $V_{in}^A = V_{in}^B$ and the sampled modulating signal is fully recovered in Bob. As shown in Figs. 5 and 6 the synchronism is not perfect all the time, however, it is good enough to make the cosine in (31) close to one. It can be seen in Fig. 13 the result of a simulation of the transmission of $\sin(kT/1000)$ ($R|\alpha|^2 mA/2 = 1$) with the presence of noise due to the non-perfect synchronism.

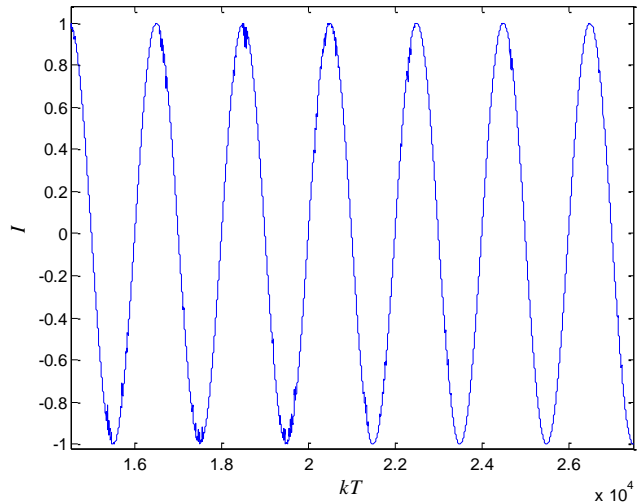

Fig. 13 – $I$ versus $kT$.

Hence, as can be seen in Fig. 13, it is enough to pass the current $I$ by a low-pass filter in order to recover the modulating signal.

## 5. Conclusions

Based on the strategy to synchronize two OEOs, we presented a strategy to synchronize three OEOs working in chaotic regime. The synchronization was observed numerically, showing transitions between perfect and non-perfect synchronization due to parameter mismatch noise. Following, we explained two applications of the synchronized OEOs in secure communication. In the first case, each local OEO works as a pseudo-random number generator. The discretization of the Stokes' parameter $S_1$ of the output field was used to produce a binary sequence that can be used as a key in cryptographic protocols. The errors in the keys obtained by Alice and Bob are due to the non-perfect synchronism and the presence of external noise. Error rate of up to 15% is acceptable since an error correction protocol can be used for reconciliation of the keys in Alice, Bob and Charlie. As expected, the BER is larger in the scheme with three synchronized OEOs. However, even in this case a useful key can be generated if the noise strength is moderated ($N > 50$).

In the second application, it was shown an optical setup for secure transmission of a sampled analog signal. Once more, the better the synchronization the better is the fidelity of the recovered signal. The setup uses phase modulation hence several details must be taken into account in order to guarantee a good interference between Alice's and Bob's signals. For example, because of channel's loss, Bob has to attenuate his signal in the same amount that the optical field sent by Alice experimented during fibre propagation. The effects of channel propagation and noise in the detectors will be addressed in a future work.

Since the synchronization of OEOs working in the chaotic regime requires two-way communication, the main disadvantage of secure communication using synchronized OEO's is the fact that the transmission rate depends on the distance between the OEOs. In other words, the time separation between consecutive pulses emitted by the laser, $\tau$, is (roughly) equal to the time required by the information produced in one OEO to arrive at the other, $L/V_g$, where $L$ is the distance between two OEOs and $V_g$ is the group velocity. Hence, high rates are possible only for short distances. For a different reason, this transmission rate versus distance behaviour also happens in quantum key distribution: the larger the distance, the larger the loss and the lower the transmission rate [24]. At last, one can note in Figs. 3 and 7 that the information sent from one OEO to another OEO is electrical. However, this does not mean that necessarily they have to exchange electrical signals. Instead, they can exchange the optical field $E_6$. For example, in Fig. 3, Alice (Bob) sends half of her (his) $E_6$ to Bob (Alice) via an optical fiber and Bob (Alice) measures it. Another possibility to have a fully optical synchronization would be Alice (Bob) to send her (his) $E_6$ to Bob (Alice) via an optical fiber and Bob (Alice) to guide it to the second PBS input. This configuration will be addressed in future works.

## Acknowledgments

This work was supported by the Brazilian agency CNPq via Grant no. 307062/2014-7. Also, this work was performed as part of the Brazilian National Institute of Science and Technology for Quantum Information.